# Defects as a factor limiting carrier mobility in WSe$_2$: a spectroscopic investigation

Zhangting Wu,[1] Zhongzhong Luo,[2] Yuting Shen,[3] Weiwei Zhao,[4] Wenhui Wang,[1] Haiyan Nan,[1] Xitao Guo,[1] Litao Sun,[3] Xinran Wang,[2] Yumeng You,[5,*] and Zhenhua Ni[1,*]

[1]*Department of Physics and Jiangsu Key laboratory for Advanced Metallic Materials, Southeast University, Nanjing 211189, China.*

[2]*National Laboratory of Solid State Microstructure, School of Electronic Science and Engineering, National Center of Microstructures and Quantum Manipulation, Nanjing University, Nanjing 210093, China.*

[3]*SEU-FEI Nano-Pico Center, Key Laboratory of MEMS of the Ministry of Education, Southeast University, Nanjing 210096, China.*

[4]*Jiangsu Key Laboratory for Design and Fabrication of Micro-Nano Biomedical Instruments, School of mechanical engineering, Southeast University, Nanjing 211189, China.*

[5]*Ordered Matter Science Research Center, Southeast University, Nanjing 211189, China.*

*Corresponding authors: zhni@seu.edu.cn and youyumeng@seu.edu.cn

## ABSTRACT

The electrical performance of two dimensional transitional metal dichalcogenides (TMDs) is strongly influenced by the amount of structural defects inside. In this work, we provide an optical spectroscopic characterization approach to correlate the amount of structural defects and the electrical performance of WSe$_2$ devices. Low temperature photoluminescence (PL) spectra of electron beam lithography (EBL) processed WSe$_2$ presents a clear defect-induced PL emission due to excitons bound to defects, which would strongly degrade the electrical performance. By adopting an e-beam-free transfer-electrode technique, we are able to prepare backgated WSe$_2$ device with limited amount of defects. A maximum hole-mobility of about 200 cm$^2$/Vs was achieved due to reduced scattering sources, which is the highest reported value among its type. This work would not only provide a versatile and nondestructive method to monitor the defects in TMDs, but also a new route to approach the room temperature phonon-limited mobility in high performance TMDs devices.

# 1 Introduction

Transitional metal dichalcogenides (TMDs) have attracted great attentions and are considered as promising candidates for next-generation electronics and optoelectronics [1-4]. Both high on/off ratio in logic transistor and high gain of photoresponse have been achieved [5, 6]. At the same time, the carrier mobility in TMDs devices is still limited by structural defects [7], charged impurities [8] and other types of traps[9,10], which in turn restrict its electronic and optoelectronic performance. Direct evidence by high resolution transmission electron microscope (TEM) has revealed that structural defects (sulphur vacancies) exist in intrinsic $MoS_2$, which introduce localized donor states inside the bandgap and result in hopping transport at room temperature [9]. The vacancies can be partially repaired by chemical functionalization of sulphur containing groups in order to improve the mobility ($\mu$) [11]. Meanwhile, defects including vacancies could also be introduced externally during the device fabrication process, e.g. the sample transfer process [12], and electron beam irradiation during the routine electron beam lithography (EBL) process.

To thoroughly study the structural defects, modern characterization techniques have been employed, such as TEM [7, 13, 14], scanning tunneling microscope (STM) [15, 16], and X-ray photoelectron spectroscopy (XPS) [17]. Although TEM and STM provide structural image in the atomic scale, they have the disadvantages of complicated sample preparation and small inspection areas [7, 13-16]. While a statistic method such as XPS suffers from its poor spatial resolution [17]. On the other hand, optical spectroscopic methods such as Raman and photoluminescence (PL) spectroscopies are also helpful for such applications because they are time efficient and nondestructive [18, 19]. Furthermore, by manipulating the focused laser beam, such optical methods can have spatial resolution varying from sub-micrometer to centimeter [20, 21]. Previously, Raman spectroscopy has been demonstrated as an efficient tool on defect characterization in graphene [22, 23]. However, the defect induced D peak in graphene is strictly limited to graphitic materials according to the double resonant Raman process [24]. Raman spectra of defective $MoS_2$ by ion bombardments indeed show that they are rather insensitive to the amount of defects [18]. At same time, in low temperature photoluminescence (PL) measurement, a defect related PL peak could be observed due to the emission of excitons bound to the defect sites [19]. However, the relation of such bound exciton emission with the electrical performance is still uncertain. Therefore, a systematic optical spectroscopic investigation of defects and their correlation with the electrical performance is highly desirable.

In this work, we clearly observed the defect induced PL emission on EBL processed WSe$_2$, as compared to pristine sample. With precisely controlling the defect density, we accomplished an optical spectroscopic characterization approach to correlate the amount of structural defects and the electrical performance of WSe$_2$ devices. It is found that $1/\mu$ is linearly dependent on the relative intensity of the defect-induced bound exciton emission. Furthermore, by adopting an e-beam-free technique, we are able to fabricate backgated WSe$_2$ devices with extraordinary hole mobility of ~200 cm$^2$/Vs, which is the highest value among its type. These results would therefore not only provide a versatile tool to monitor the defects in TMDs but also introduce a new route to fabricate high mobility TMDs sample for electronic and optoelectronic applications.

**2 Experimental**

**2.1 Preparation of monolayer WSe$_2$ and optical characterization.**

Monolayer WSe$_2$ flakes were exfoliated from bulk crystals (SPI Supplies) and transferred to Si wafer with a 300 nm SiO$_2$ capping layer. The thickness of monolayer WSe$_2$ is confirmed by Raman spectra and optical contrast [25]. The Raman and PL spectra were recorded using a LabRAM HR800 Raman system with excitation of 514.5 nm laser light and a 50x objective lens with numerical aperture of 0.5. Low temperature measurements were performed in an INSTEC HCP621V stage with mk1000 high precision temperature controller and LN2-SYS liquid nitrogen cooling system.

**2.2 Device fabrication and electrical measurement.**

Two techniques were adopted to fabricate the WSe$_2$ FET. One is that we used standard EBL to pattern the electrodes of WSe$_2$ backgated FETs on 300nm SiO$_2$/Si substrates, as also shown in Fig. 1. Au (50nm) was thermal evaporated as the contact metal for source, drain and voltage probes. Another one was fabricated by transferring Au electrode, an e-beam-free fabrication process. The 100 nm thick Au electrodes (with size of 150×300 $\mu$m) were prepared by thermal evaporating on SiO$_2$/Si substrates. We used a tungsten probe tip attached to a micro manipulator to carefully pick up the Au electrodes and transfer onto monolayer WSe$_2$ sample under microscope [26]. All devices, including e-beam-free and EBL devices, were annealed at 130 °C in a vacuum chamber with a gas mixture of H$_2$ and Ar (with ratio of 20/100 and pressure of 100 Pa) for two hours to improve contacts [27]. Electrical measurements were carried out by a Keithley 2612 analyzer under ambient environment. The devices exhibit very small hysteresis at room temperature.

## 3 Results and discussion

### 3.1 Defect induced PL emission

With regards to the nanofabrication process, EBL is the most commonly used technique to fabricate microelectronic devices [28]. However, during the EBL process, electron beam would unavoidably irradiate on TMDs through the top PMMA layer. Fig. 1a and inset of Fig. 1b show the schematic and optical image of a monolayer WSe$_2$ flake with a top PMMA layer patterned by EBL. The electron beam (20kV accelerating voltage and $7.5\times10^6$ $\mu m^{-2}$ electron density) irradiated PMMA region has been removed by chemicals which appears pink in the optical image, due to the optical interference of 300 nm SiO$_2$/Si substrate. The PL spectra of pristine and electron beam irradiated WSe$_2$ measured at low temperature of 83K are shown in Fig. 1b, respectively. The pristine monolayer WSe$_2$ exhibits two PL peaks located at ~1.71 and ~1.73 eV at 83 K, which have been intensively investigated and assigned to the emissions from trions and excitons, respectively [29]. On the other hand, the EBL sample presents an additional PL peak located at ~1.62 eV (named as X$_b$ peak here). Similar low temperature feature has also been observed previously in defective MoS$_2$ and WSe$_2$ samples, and was attributed to the exciton bound to defects or absorbed impurities [19,30], which would introduce mid-bandgap states with energies ~0.1-0.3 eV above the valence band maximum or below the conduction band minimum as calculated by theory [19,31]. A control experiment was done on another piece of exfoliated sample, showing that coating and removing PMMA layer would not contribute to the appearance of X$_b$ peak, as shown in the Electronic supplementary material Fig. S1a. There is also a concern that the deposition of amorphous carbon would happen in the SEM chamber which acts as absorbed impurities contributing to the X$_b$ peak. We have to mention that the PMMA layer of uncovered region was removed after the EBL process, and such possibility can be excluded. More data on the control experiment is shown in the Electronic supplementary material Fig.S1b and Fig. S2. Therefore, the appearance of $X_b$ peak in EBL processed WSe$_2$ is not related to charged impurities or other adsorbate, but due to the structural defects introduced by e-beam irradiation, such as single vacancies, di-vacancies, cluster of vacancies, etc. [7,13].

### 3.2 Bound exciton and activation energy

For a detailed study, we investigated the temperature dependence of $X_b$ peak, and the results are shown in Fig. 2a and b. To avoid the laser caused damage, the incident power was controlled at a moderated level (0.6 $\mu$W). As can be seen in Fig. 2a, the PL intensity of $X_b$ peak is vanished with increasing temperature, which is understandable since excitons are not tightly bound to defects and such weak interaction can be easily perturbed by thermal stimulation [32]. On the other hand, the increase of PL intensity from neutral excitons (names as $X_0$ here) with the increase of temperature has been well explained by Zhang, *et. al.* as the effect from dark excitons [33]. Therefore, the intensity ratio between $X_b$ and $X_0$ decrease dramatically with the increase of temperature. For steady-state condition, we expect that the observed variation in $X_b$ intensity with temperature is predominantly a reflection of the $X_b$ population, and reveals the characteristic of a full thermal equilibrium between bound excitons and free excitons. A thermally dissociation process can be used to describe the population of $X_b$, that is $N_{Xb}$, as a function of temperature T: $N_{xb}(T) = \dfrac{N_0}{1+(\tau/\tau_0)e^{-E_A/kT}}$, where $\tau$ is the excitonic lifetime exceeding 100 ps, $\tau_0$ is effective scattering time and $E_A$ is activation energy [32, 34]. The thermal activation energy here describes the necessary energy of thermal perturbation which prevents the free excitons from trapping by the localized defect sites. By fitting the data in Fig. 2b with the Equation, we get $E_A$ of 43 meV and the ratio $\tau/\tau_0$ of 259 for bound exciton in monolayer WSe$_2$.

The population of bound exciton $N_{Xb}$ is not only affected by temperature, but also the number of excitation photons. The PL spectra at different excitation power are shown in Fig. 2c. Since the intensity of $X_0$ at such a low excitation density (less than 1.5 $\mu$W) can be treated as linear dependence on incident laser power [34], we normalized the PL spectra by $X_0$ intensity to emphasize the change of $X_b$ intensity. By extract the peak area (I) of $X_b$ with excitation power (P), as plotted in Fig. 2d, an obvious sublinear dependence can be observed and can be well fitted by power law $I \propto P^k$ [35], where k is ~0.59, whereas the free exciton intensity ($X_0$) scales linearly. Such a nonlinear laser power dependence and tendency to saturate can be explained by the full population of defect states with excitons at high excitation power. Therefore, the saturated PL intensity of $X_b$ could be used to effectively estimate the defect concentration in WSe$_2$.

**3.3 Defects limited carrier mobility in WSe$_2$ devices**

Next, we investigated the effects of defects as probed by optical spectroscopic measurements on the electrical performance of WSe$_2$. For comparison, we fabricated two monolayer WSe$_2$ devices by EBL and e-beam-free electrode-transfer technique, respectively, and their transport properties are shown in Fig. 3. As can be seen, both devices exhibit *p*-type transistor behaviors, while the device prepared by e-beam-free technique (L = 6.1 $\mu m$, W = 5.1 $\mu m$) has a significantly higher field-effect mobility $\mu = \frac{\partial I_{DS}}{\partial V_{GS}} \frac{L}{WC_g V_{DS}}$ of ~200 cm$^2$/Vs at $V_{GS}$ = -70V and on/off ratio of ~10$^6$ as compared to the EBL device, which has a mobility of only ~0.4 cm$^2$/Vs at $V_{GS}$=-70V and on/off ratio of only ~10$^2$. We did not observe the ambipolar behavior as previous reported in some literatures, the reason may be the different work functions between Au and other metals such as Ti [5, 40]. We have prepared around ten sets of devices and all the EBL devices present mobility <1 cm$^2$/Vs, and the e-beam free devices have mobility in the range of 20-200cm$^2$/Vs, due to different contact between transferred metal electrodes and WSe$_2$. The statistical results of mobility are shown in Electronic supplementary material Fig. S3. The great improvements of carrier mobility by more than 100 times and on/off ratio by ~10$^4$ indicate that charge injection into the channel is more efficient and device performance is dramatically enhanced in the e-beam-free device. As both devices were tested under the same conditions, by eliminating the effect of chemicals involved (see Electronic supplementary material in Fig.S1.), such a significant difference should be caused by e-beam irradiation induced defects, as revealed by the strong $X_b$ peak in EBL processed WSe$_2$. It has been demonstrated that the electrical transport in MoS$_2$ and other TMDs is strongly influenced by extrinsic trapping sites and scattering centers such as charged impurities and structural imperfection, leading to a much lower mobility than the calculated intrinsic limit [9, 11, 36]. Comparing to WSe$_2$, the density of defects in intrinsic MoS$_2$ sample is considerably large [9], which is also confirmed by the intensive $X_b$ peak for pristine MoS$_2$ sample, as shown in electronic supplementary material Fig. S4. The defects associated localized states form in the bandgap of MoS$_2$, which appear as electron donors and result in hopping transport behavior below a critical carrier density [9]. At same time, the defects are also efficient scatters for free carriers, reduce their lifetime and hence the carrier mobility. Same mechanisms applied in our $WSe_2$ devices and the poor mobility and on/off ratio in the EBL sample can be well explained due to defects introduced by e-beam irradiation. To minimize the effect from defects, approach had been made by low-temperature thiol chemistry route to repair the sulfur vacancies and mobility >80 $cm^2V^{-1}s^{-1}$ has been achieved in backgated monolayer $MoS_2$ FET at room temperature [11]. To improve the performance of $WSe_2$ devices, there are works focusing on improving contact between metal and semiconductor [5, 37-39], adopting high-k top-gate insulator [40], and a maximum hole mobility of ~100 $cm^2V^{-1}s^{-1}$ in backgated device had been demonstrated [39]. Here, by using a simple transfer-electrode technique, we avoided e-beam irradiation damages and achieved the highest reported room-temperature mobility of ~200 $cm^2$/Vs in backgated $WSe_2$ samples. It should be noted that, there is still plenty of room for further improvement such as improving contact between $WSe_2$ and transferred electrode [5, 37-39], reducing charged impurities on substrate [41] and minimizing adsorbed molecules on $WSe_2$ surface [40]. By carefully considering the above issues, the room temperature phonon limited electron mobility of 700 $cm^2$/Vs can hopefully be approached [42].

**3.4 Controllable introduction of defects and correlation with PL and mobility**

Since the dosage of e-beam can be precisely controlled, we can then fine-tune the density of defects in $WSe_2$ and correlate it with the device performance. Fig. 4a shows the evolution of PL spectra of monolayer $WSe_2$ under different irradiation electron density. In order to slow down the progress of defect introduction, the accelerating voltage of e-beam was kept at 9 kV. As can be seen, while pristine $WSe_2$ presenting negligible $X_b$ peak, in the e-beam irradiated sample, $X_b$ peak with finite intensity can be observed and its intensity increases monotonically by increasing electron dosage. Another feature is that Raman $A_{1g}$ mode and $E_{2g}$ mode of the sample remain almost unchanged after e-beam irradiation, as shown in electronic supplementary material Fig. S5. This indicates that Raman spectroscopy is insensitive for detecting the relatively small amount of defects in TMDs, since it only probes the vibrational properties and the perturbation in crystal lattice [18]. On the other hand, the PL intensity of neutral excitons $X_0$ peak gradually decreases with the increase of electron irradiation density, as shown in Fig. 4a. This is because the emission process of $X_0$ is determined by the exciton lifetime[32], which is greatly reduced by the introduction of defects and hence more relaxation channels. Therefore, PL is more sensitive tool to study defects in TMDs as compared to Raman.

To quantify the defect density with different electron dosage, samples with electron irradiation density of 7.28, 14.47 and 37.97 ×10$^6$ $\mu m^{-2}$ were chosen and the X$_b$ intensities were plotted against laser power as shown in Fig. 4b. Following above discussed method, a power-law fitting was applied and similar k values of ~0.6 were obtained, which suggest that the density of e-beam does not alter the nature of induced defects. The saturated intensity of X$_b$ rises with increasing e-beam density, in good accordance with the increase of defect densities at higher dosage. Fig. 4c shows the ratio of integrated peak area between X$_b$ and X$_0$ taken at the excitation laser power of 1.45 $\mu$W. Here, the intensity of X$_0$ peak was used for normalization as the intensity of X$_b$ peak is affected by different experimental conditions, similar to defect characterization in graphene by Raman intensity ratio of D and G peaks (I$_D$/I$_G$) [23]. It can be seen that such ratio shows very good linear dependence with the irradiation electron dosage in the range of <60 ×10$^6$ $\mu m^{-2}$, which suggest that the low temperature X$_b$ peak can be used as a standard approach to characterize and monitor the defects in WSe$_2$ sample. Since the emission mechanism of such bound exciton is similar in TMDs, this method can be well applied over all TMD family, e.g. MoS$_2$, WS$_2$, etc. When irradiation electron density is larger than 60 ×10$^6$ $\mu m^{-2}$, the intensity ratio begins to decrease, which illustrates that WSe$_2$ begins to be seriously damaged. Under such a high electron dosage, the WSe$_2$ becomes a disordered system, and in that case, the exciton behavior cannot be described by currently adapted two-dimensional approximation, thus falling out of our scope of discussion. Although there is a good correlation between the ratio I$_{Xb}$/I$_{X0}$ and e-beam dosage, one of the remaining concerns is that we are still unable to measure the exact number of defects in EBL processed WSe$_2$, e.g. by using high resolution TEM, since TEM itself would provide more defects in WSe$_2$ due to e-beam irradiation. Nevertheless, this again suggests that optical method has great advantage on the identification of defects for two dimensional materials, especially on e-beam sensitive materials, like TMDs.

Furthermore, owing to the advantage of non-destructivity, our optical method allows direct correlation between X$_b$ intensity and the device performance, and provides a pronouncing practical application potential. In this purpose, we studied the electrical properties of WSe$_2$ devices as a function of defect density (represented by e-beam dosage) and bridge them by the ratio $I_{Xb}/I_{X0}$. The transfer curves of a pristine WSe$_2$ device prepared by transfer-electrode method under different irradiation electron density are shown in Fig. 5a. Due to the protection of Au electrodes, the electron beam only irradiates at the channel of device. The mobility of this pristine WSe$_2$ device is 150cm$^2$V$^{-1}$s$^{-1}$. This mobility decays dramatically with the increase of electron irradiation density. After the irradiation density of $12.9\times10^6$ μm$^{-2}$, the mobility is only about 13 cm$^2$V$^{-1}$s$^{-1}$, decreases by more than 10 times compared to the pristine value, as shown in inset of Fig. 5b. These results show that the small amount of defects introduced by electron irradiation can produce a great influence on electric properties of WSe$_2$ device. It should be noted that the intensity of $X_b$ peak of EBL processed WSe$_2$ is much stronger (Fig. 1b), which results in even lower mobility for the EBL device as shown in Fig. 3. According to Matthiessen's rule [11], the mobility for free carriers is expressed as:

$$\mu_0(n,T)^{-1} = \mu_{ph}(T)^{-1} + \mu_{CI}(n,T)^{-1} + \mu_{sr}^{-1},$$

where $\mu_{ph}, \mu_{CI}, \mu_{sr}$ are mobility limited by phonons (ph), charged impurities (CI) and short-range (sr) scatterings, respectively. The defects, e.g. vacancies, are efficient short-range scatters for conducting carriers, which would result in $1/\mu \propto N_{sr}$, where N$_{sr}$ is the number of scattering centers or defects and presented by the electron irradiation dosage in this study. Fig. 5b does show a linear increase of $1/\mu$ with the increase of the electron irradiation dosage, and also the intensity ratio between $X_b$ and $X_0$ ($I_{Xb}/I_{X0}$). The above results provide appealing evidence that defects responsible for the appearance of $X_b$ PL peak can also set up a limit on the carrier mobility of WSe$_2$. Other factors such as phonon and impurities scatterings could contribute to the offset in the $y$ axis at $I_{Xb}/I_{X0}$ ~0 in Fig. 5b.

**4 Conclusions**

In summary, we have shown that structural defects could be introduced in WSe$_2$ device even in the routine EBL process, and greatly degrade the device performance. By carrying out low temperature PL spectroscopic analysis and transport study, we successfully correlated the carrier mobility with the relative intensity of defect related $X_b$ peak in PL spectrum. Furthermore, by adopting an e-beam-free transfer-electrode technique, we were able to prepare WSe$_2$ device with limited numbers of defects and the electrical performance has been greatly improved, and a maximum hole-mobility of about 200 cm$^2$V$^{-1}$s$^{-1}$ was achieved due to reduced scattering sources. This work would therefore not only provide a nondestructive and efficient method to investigate the defects, but also an approach to the ultimate goal of fabricating defect-free device with intrinsic electrical performance of TMDs material.

## Acknowledgements

This work was supported by NSFC (61422503 and 61376104), Jiangsu key laboratory for advanced metallic materials (BM2007204), the open research funds of Key Laboratory of MEMS of Ministry of Education (SEU, China), and the Fundamental Research Funds for the Central Universities. The authors would like to thank Prof Zhenhua Qiao from USTC, China for helpful discussions.

**FIGURES.**

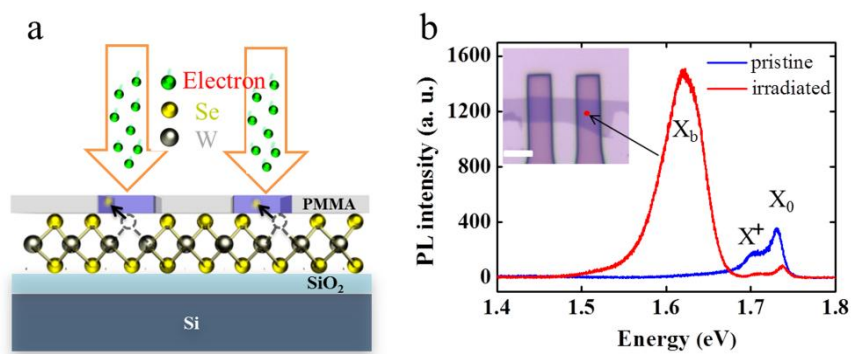

**Figure 1** Defect induced PL emission. (a) Schematic diagram of electron beam irradiation on monolayer $WSe_2$ sample during the EBL process. The blue PMMA regions will be washed by developer (MIBK: IPA, 1:3) and stopper (IPA, 2-Propanol) after the EBL process. (b) PL spectrum of pristine monolayer $WSe_2$ and monolayer $WSe_2$ after EBL with 20kV accelerating voltage and $7.5 \times 10^6$ $\mu m^{-2}$ electron density. The inset shows optical image of $WSe_2$ with PMMA patterned by EBL, scale bar is 5 $\mu m$.

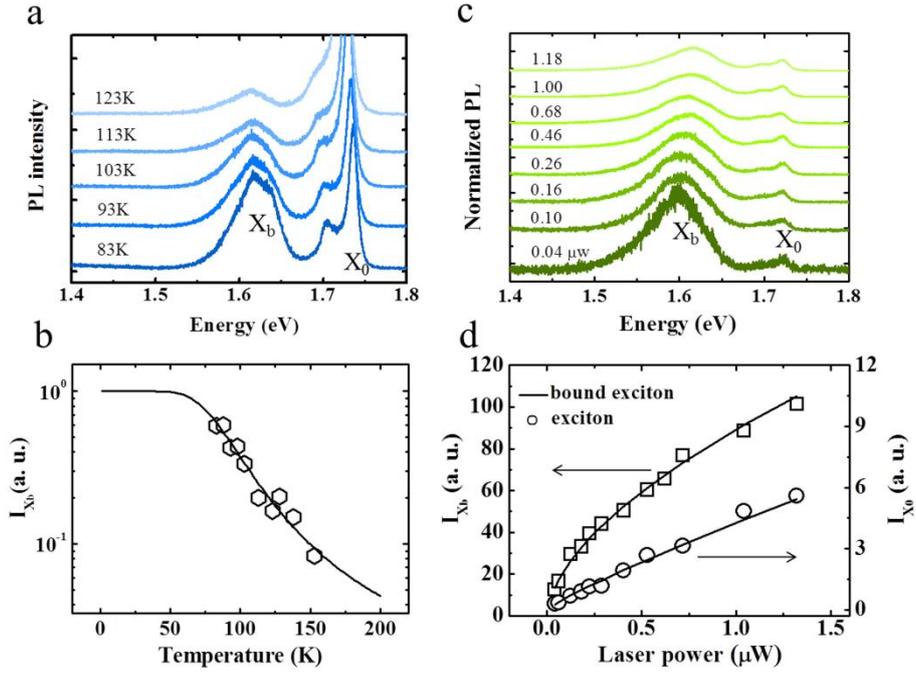

**Figure 2** PL characteristics as a function of temperature and laser power. (a, b) Temperature dependence of PL spectra and $X_b$ intensity. The solid curve in (b) is the fitting result based on the thermal dissociation model described in the text, which is related to the activation energy $E_A$ and effective scattering time $\tau_0$. (c) PL spectra under different laser power and normalized by the intensity of $X_0$. (d) The laser power dependence of integrated PL intensity of $X_b$ and $X_0$. The solid curve is the fitting result according to power law $I \propto P^k$ described in the text.

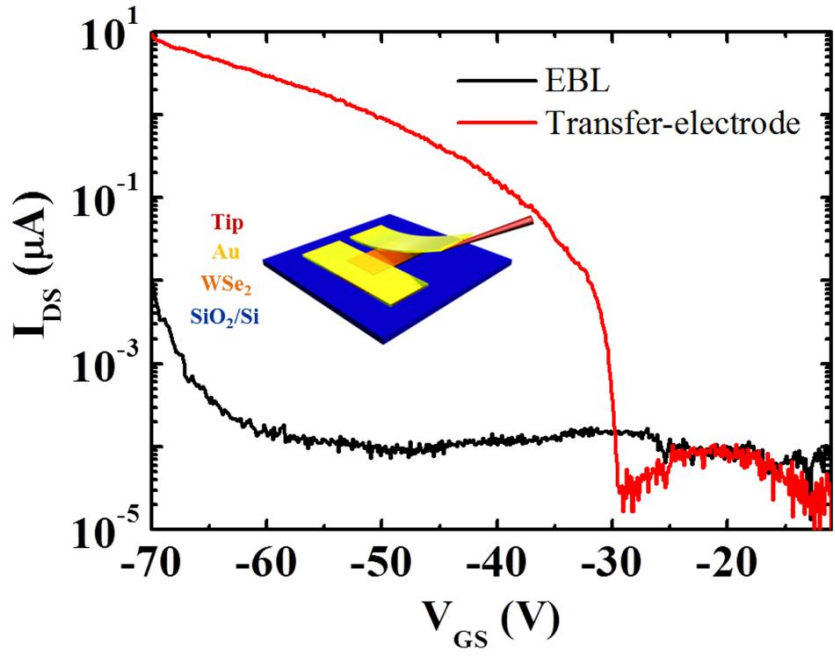

**Figure 3** Electrical properties of EBL and e-beam-free WSe$_2$ devices. Transfer curves of current ($I_{DS}$) versus gate voltage ($V_{GS}$) of two WSe$_2$ devices prepared by EBL (black curve) and e-beam-free transfer-electrode (red curve) techniques at $V_{DS}$=0.4V. The inset shows schematic diagram of transferring Au electrodes on WSe$_2$ sample.

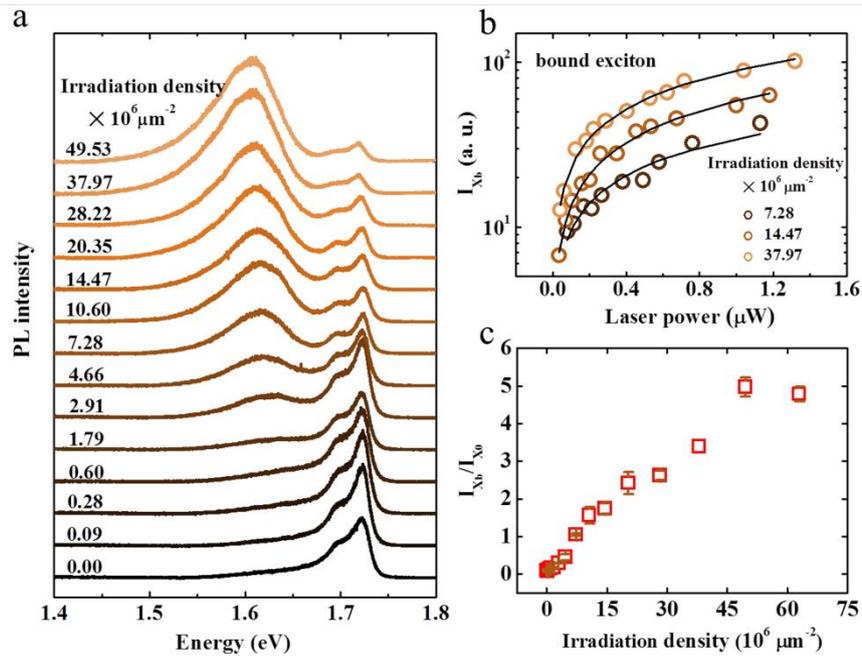

**Figure 4** PL characteristics as a function of electron beam irradiation. (a) PL spectra of a pristine WSe$_2$ under different e-beam irradiation density. (b) The excitation power dependence of the integrated intensities of X$_b$ peak with electron density 7.28, 14.47, and 37.97 $\times 10^6$ $\mu m^{-2}$. The solid curve is the fitting result according to power law $I \propto P^k$ described in the text. (c) The intensity ratio of X$_b$ and X$_0$ with increasing irradiation electron density. The accelerating voltage in SEM is 9kV.

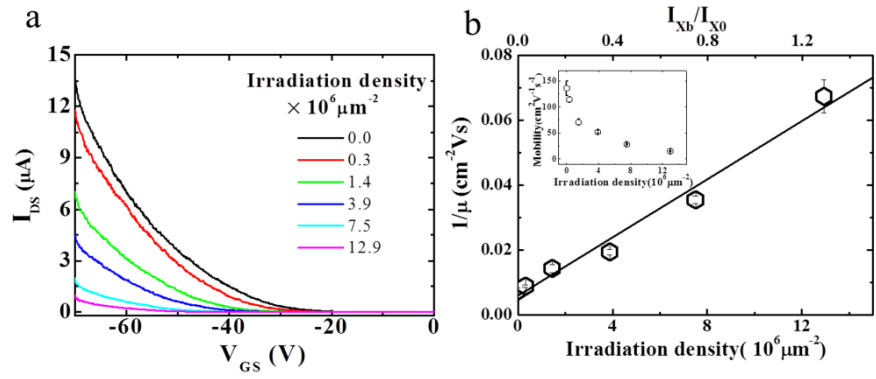

**Figure 5** Transport characteristics as a function of electron beam irradiation. (a) The transfer curves $I_{DS}$-$V_{GS}$ of a WSe$_2$ device after e-beam irradiation with different dosage. (b) Changes of scattering rate $1/\mu$ and mobility $\mu$ (inset) as a function of e-beam dosage, and the intensity ratio of $X_b$ and $X_0$.

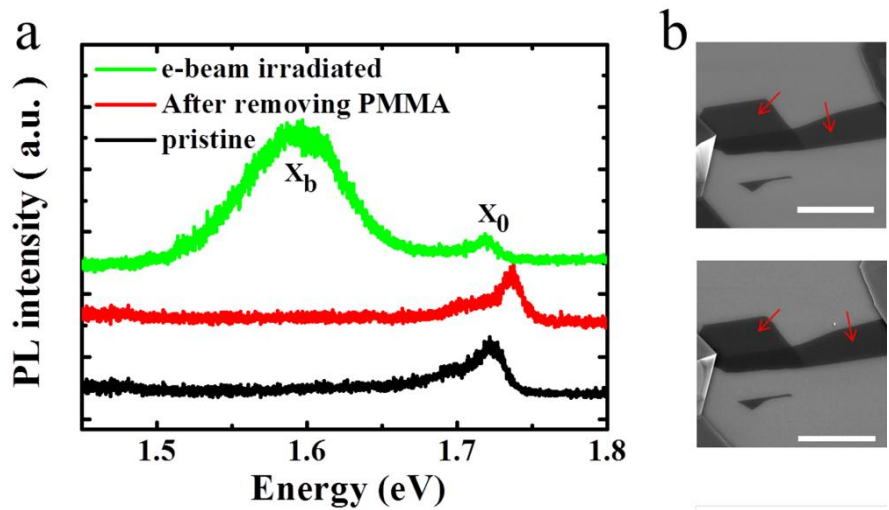

**Figure S1** (a) PL spectra of pristine, coating and removing PMMA, and e-beam irradiated monolayer $WSe_2$ samples taken at 83K with 0.04 $\mu$W laser power. $X_b$ peak is absent in the PL spectrum of PMMA removed monolayer $WSe_2$, which confirms that PMMA residues or other adsorbate would not contribute to the appearance of $X_b$ peak. (b) The SEM images of single layer $WSe_2$ sample before coating (up) and after removing (down) PMMA layer. Scale bar is 10 $\mu m$. The surface of the sample is very smooth, which indicates that PMMA is removed.

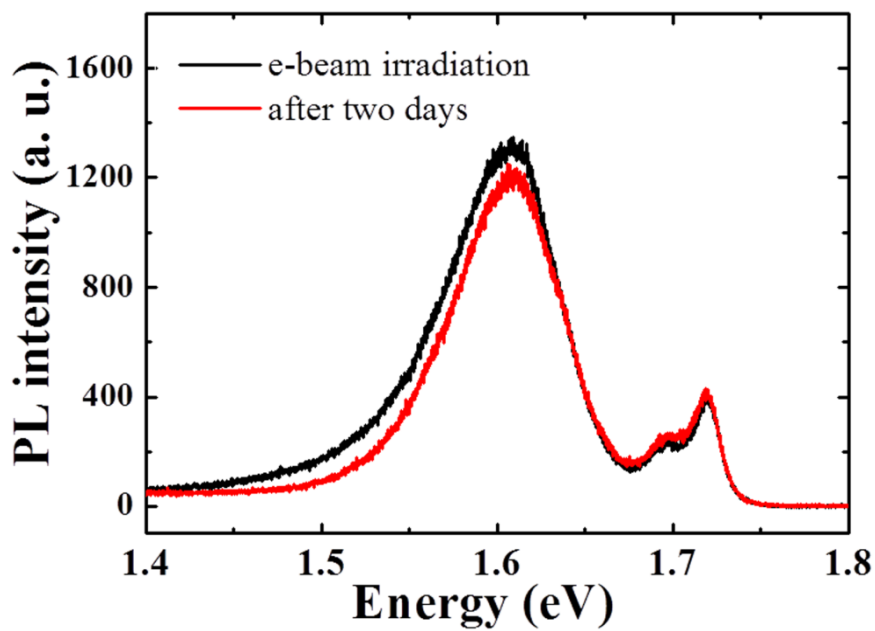

**Figure S2** PL spectra of e-beam irradiated $WSe_2$ sample taken at 83K with 1.10 $\mu$W laser power. The black one is measured right after e-beam irradiation. The red one is measured again after two days. No obvious difference has been observed.

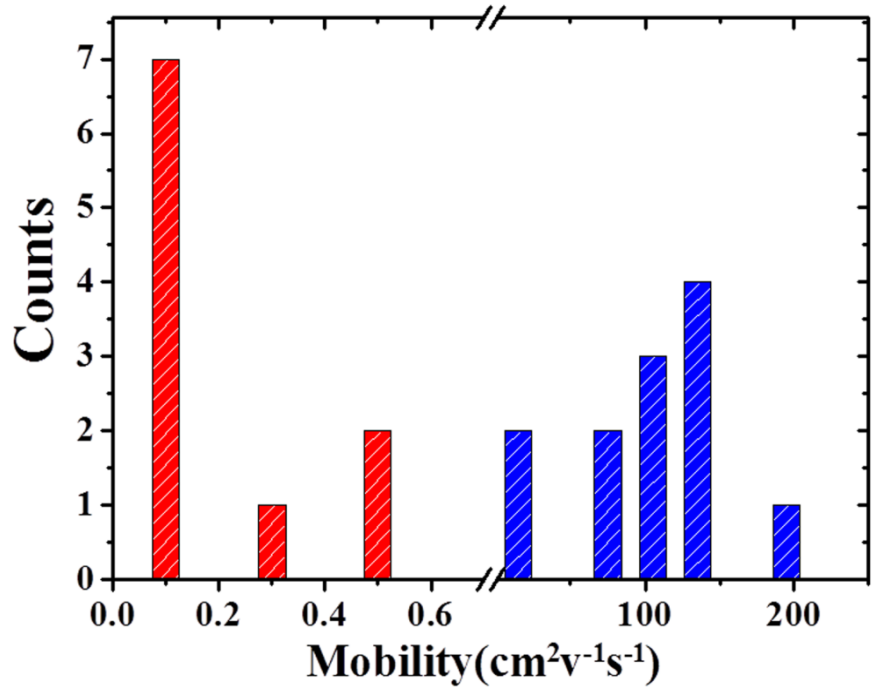

**Figure S3** Statistics of hole mobility of WSe$_2$ for EBL (red) and e-beam free devices (blue).

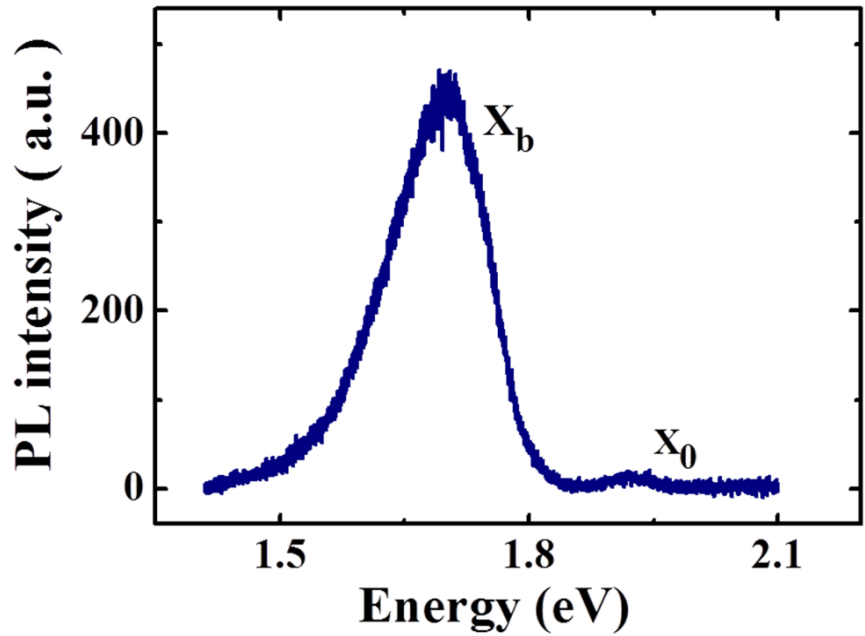

**Figure S4** PL spectrum of mechanically exfoliated monolayer MoS$_2$ taken at 83K with 1.4 $\mu$W laser power. An intensive X$_b$ peak is presented, indicating that the density of defects in MoS$_2$ sample is considerably large.

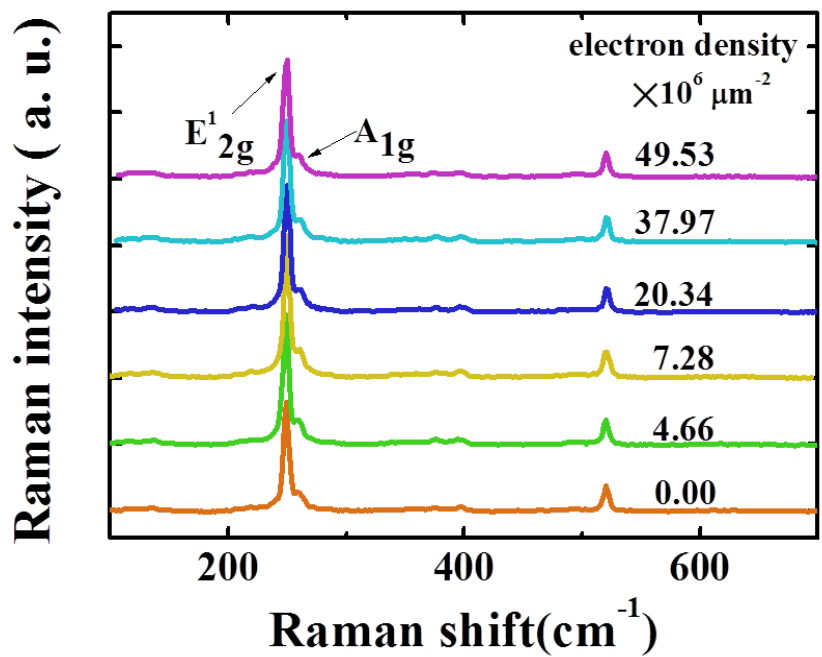

**Figure S5** Raman spectra of a pristine WSe$_2$ under different electron beam irradiation densities. Raman A$_{1g}$ mode and E$_{2g}$ mode of the sample remain almost unchanged with the increase of e-beam irradiation and no new peaks are presented, indicating that Raman spectroscopy is insensitive for detecting the amount of defects introduced by e-beam irradiation.